\documentclass[aps,prb,twocolumn,groupedaddress,showpacs]{revtex4-1}
\usepackage{color}
\usepackage{amsmath,amsthm,amssymb}
\usepackage{graphics}
\usepackage{graphicx}
\usepackage{epsfig}
\usepackage{hyperref}
\usepackage{subfigure}
\usepackage{float}

\begin{document}

\title{Interplay of Superconductivity and Spin-Dependent Disorder}

\author{R.~Nanguneri$^1$, M.~Jiang$^{1,2}$, T. Cary$^3$,
  G.G.~Batrouni$^{4,5,6}$, and R.T.~Scalettar$^1$}

\affiliation{$^1$Physics Department, University of California, Davis,
  California 95616, USA}
\affiliation{$^2$Department of Mathematics, University of California, Davis,
  California 95616, USA}
\affiliation{$^3$Department of Physics, Rensselaer Polytechnic
  Institute, Troy, New York 12180-3590, USA}
\affiliation{$^4$INLN, Universit\'e de Nice-Sophia Antipolis, CNRS;
  1361 route des Lucioles, 06560 Valbonne, France}
\affiliation{$^5$Institut Universitaire de France}
\affiliation{$^6$Centre for Quantum Technologies,
National University of Singapore; 2 Science Drive 3, Singapore 117542}

\begin{abstract}
The finite temperature phase diagram for the 2D attractive fermion
Hubbard model with spin-dependent disorder is considered within
Bogoliubov-de Gennes mean field theory. Three types of disorder are
studied.  In the first, only one species is coupled to a random site
energy; in the second, the two species both move in random site energy
landscapes which are of the same amplitude, but different
realizations; and finally, in the third, the disorder is in the
hopping rather than the site energy.  For all three cases we find
that, unlike the case of spin-symmetric randomness, where the energy
gap and average order parameter do not vanish as the disorder strength
increases, a critical disorder strength exists separating distinct
phases.  In fact, the energy gap and the average order parameter
vanish at distinct transitions, $V_{c}^{\rm gap}$ and $V_{c}^{\rm
op}$, allowing for a gapless superconducting (gSC) phase.  The gSC
phase becomes smaller with increasing temperature, until it vanishes
at a temperature $T^{\ast}$.
\end{abstract}

\pacs{74.25.Dw, 74.62.En}
\maketitle

\noindent
\section{Introduction}
The study of disordered electronic systems is of major interest in
condensed matter physics \cite{lee85,belitz94} because randomness breaks
translational invariance and leads to localized electron states.  In
noninteracting systems, typically all the states are localized in one
and two dimensions, even by small amounts of disorder \cite{abrahams79},
while in three dimensions, states with large and small energies are
localized, and a `mobility edge' separates these from extended states
near the center of the energy distribution.  In real materials, disorder
can arise from vacancies or impurity atoms, dislocations, and other
forms of structural imperfections.

The effect of interparticle interactions on these localized and extended
states has proven to be a very challenging problem.  Experiments using
silicon MOSFETS of very high purity now suggest
\cite{kravchenko94,popovic97,abrahams00} that repulsive interactions
might allow a metallic phase to exist in two dimensions.  An equally
interesting set of questions concerns the interplay of disorder and
attractive interactions, and the occurrence of superconductor-insulator
transitions.  One well-studied issue is the existence of a universal
conductivity in thin films \cite{arwhite86,orr86,fisher90,hebard94}.

Recently it has become possible to address the interplay of randomness
and interactions within a new experimental context, namely that of
ultracold fermionic and bosonic atomic gases confined to optical
lattices \cite{jaksch98,bloch08}.  Here the role of `spin' is played by
atoms in different hyperfine states.  Disorder can be superimposed on
the periodic optical lattices created by interfering counter-propagating
lasers by means of the introduction of speckle
fields\cite{aspect09,palencia10}.  This randomness can now be made on
the scale of the lattice constant.  Interactions between the atoms can
be tuned through a Feshbach resonance\cite{chin10} and can be either
repulsive or attractive.

Optical lattice experiments do not just represent a new realization
of model Hamiltonians of interacting and disordered
materials\cite{white09,zhou10}; they also allow the study of situations
which would be more difficult to achieve in the solid state.  One
possibility is ``spin''-dependent disorder in which atoms of different
hyperfine structures do not see the same random potential, a natural
follow-on to existing experiments where the optical lattice itself is
spin-dependent\cite{demarco11,demarco10,mandel03,feiguin09,liu04}.

In this paper we present a detailed theoretical study of such
spin-dependent disorder.  We consider the attractive fermion Hubbard
model in two dimensions, which has a low temperature superconducting
phase, and include three types of spin-dependent disorder: 1) random
potential on the up-spin but zero potential on the down-spin; 2) random
potential of the same strength but different realizations on each spin
species; 3) random hopping energies of the same strength, but again with
different
realizations on each spin species.. We study these cases with
finite-temperature Bogoliubov-de Gennes (BdG) mean-field
theory\cite{degennes66}. This BdG approach to the $-|U|$ Hubbard model
allows us to examine the strong interaction and strong disorder regimes
which are beyond the realm of validity of Anderson's
theorem\cite{anderson59}. The mean-field approximation of course ignores
fluctuations. Some of the effects of extending beyond the mean-field are
explored in Ref.~\onlinecite{ghosal98,ghosal01}.

Our key results are as follows: [1] Unlike the case of spin independent
randomness, for which BdG theory produces no transition as the disorder
strength is increased, we show here that spin dependence produces a
vanishing of the superconducting order parameter $\Delta_{\rm op}$ and
energy gap $E_{\rm gap}$; [2] The critical points can be different for
$\Delta_{\rm op}$ and $E_{\rm gap}$, leading to a gapless
superconducting phase; and [3] models with different types of
spin-dependent disorder behave in a qualitatively similar manner.
Results at zero temperature for this problem can also be found in
Ref.~\onlinecite{jiang11}.

We begin with a description of the model and method in Section II, and
present our most central results, the phase diagrams, in Section III.
Several additional details of the properties of these models are
summarized in Section IV.  Section V contains some concluding remarks.
The rest of this introduction will provide more details for the
experimental prospects for realizing spin-dependent disordered optical
lattices.

Spin-independent disorder in a 3D optical lattice has been studied in
the context of the disordered Bose-Hubbard model in
Refs.~\onlinecite{white09,pasienski10}.  The phase diagram was
determined through transport properties.  Non-interacting fermions in a
spin-independent disordered 3D lattice have also been
realized\cite{kondov11} and used to demonstrate Anderson localization
with an accompanying mobility edge\cite{aspect09}. However, experiments
with spin dependent disorder have yet to be performed, but seem to be
feasible \cite{demarco11,pasienski_pc11}.  Before discussing how that
might be arranged, we review how the light field producing the optical
lattice has already been coupled to atomic states in a spin-dependent
way, thereby enabling spin-dependent optical lattices of ultra-cold
atoms\cite{liu04}.

The alkali atom $^{40}K$ has a single electron in its outermost S-shell,
giving a $2S_{1/2}$ electronic configuration with spin projections
$M=\pm 1/2$ serving as the two spin-species.  $^{40}K$ also has two
excited states $2P_{1/2}$ and $2P_{3/2}$. We ignore for now the
complications of coupling between nuclear and electron spins. When the
atom is in a laser light field of frequency away from the resonant
frequency for excitation from the ground to the exited states, the
effect of the oscillating electric field is to act as a perturbation
which induces a new ground state with an electronic dipole moment by
mixing the excited states with the atomic ground state.  The induced
dipole itself interacts with the electric field. The dipole potential
energy is proportional to the product of the frequency dependent
polarization $\alpha(\omega)$ and the light intensity, $\langle
|\bf{E}(\omega)|^{2} \rangle$. By tuning the frequency of the laser
below or above resonance, the sign of $\alpha(\omega)$ can be switched.
Depending on this sign, the interaction energy is lowered if the atom is
at positions of high or low field intensity. At resonance, the atom
absorbs and emits light, which is undesirable, therefore the laser has
to be detuned away from the resonant frequency for atomic transitions.
The $|2S_{1/2},M=1/2 \rangle$ ($|2S_{1/2},M=-1/2 \rangle$) states are
coupled independently by right- (left-) circular polarizations to the
$2P_{3/2}$ ($2P_{1/2}$) excited states. By tuning the two circular
polarizations independently of each other in the optical standing-wave
lasers, a spin-dependent lattice can be realized. This optical lattice
determines the Hubbard model parameters $t$, $U$ and chemical potential
offset $\mu$.  As a consequence, these quantities can depend on the
fermionic species $\sigma$, as described above.

We now discuss how one might superimpose a disordered potential or
hopping on top of the standing wave optical lattice in a similarly spin
dependent fashion.  Present experimental techniques allow a controlled
disordered potential to be created by passing a detuned laser
(monochromatic, phase-coherent light) through a ground-glass plate
diffuser\cite{palencia10,pasienski11}. The light exits from points on
the diffuser plane with random phases and is focused by a lens at the
optical lattice located at the focal plane. The random phases interfere
to produce a Gaussian distributed electric field intensity. Since the
induced atomic dipole couples to the field intensity, the lattice
potential depth thus becomes disordered. Spin-dependent disorder can be
created by passing detuned laser light with equal components of left and
right circular polarizations through a birefringent diffuser with
thickness much greater than the wavelength, so that the two
polarizations exiting the diffuser acquire a much greater, uncorrelated
phase difference between them \cite{pasienski_pc11,demarco11}.
Alternatively, two separate non-birefringent diffusers may be used, one
for each circular polarization. In the latter scenario, it is easy to
switch off disorder on one spin species by removing one of the
diffusers. The lattice itself is created by standard techniques.

We now turn to the Hamiltonian which models such situations.

\section{Model and Method}

\subsection{Attractive Hubbard Hamiltonian}

The clean attractive Hubbard Hamiltonian
\begin{eqnarray}
H = &-t& \sum_{\langle {\bf ij} \rangle,\sigma}
(c^{\dagger}_{{\bf i}\sigma}c^{\phantom{\dagger}}_{{\bf j}\sigma} +
c^{\dagger}_{{\bf j}\sigma}c^{\phantom{\dagger}}_{{\bf i}\sigma} )
\nonumber
\\
&-& |U|\sum\limits_{\bf i}  n_{{\bf i} \uparrow}  n_{{\bf i}\downarrow}
- \mu \sum\limits_{\bf i}  (n_{{\bf i} \uparrow}  + n_{{\bf
i}\downarrow})
\label{Hex}
\end{eqnarray}
describes a set of fermions hopping with amplitude $t$ on near
neighbor sites (in this paper we consider a square lattice) and
interacting on-site with an energy $-|U|$.  The chemical potential $\mu$
controls the filling.  Away from half-filling, $\rho = \langle n_{{\bf
i} \uparrow} + n_{{\bf i}\downarrow} \rangle = 1$, there is a
Kosterlitz-Thouless transition at finite $T_c \approx 0.1 t$ to a
state with off-diagonal (superconducting) long range order
\cite{micnas81,scalettar89,denteneer94,dossantos94,paiva04}.

Within a BdG treatment the interaction term in $H$ can be decoupled in
different (charge, pairing, spin) channels.  Since $-|U|<0$ we focus on
pairing and write,
\begin{eqnarray}
H_{\rm eff} = &-& \sum_{\langle {\bf ij} \rangle,\sigma}
t_{{\bf ij}\sigma}
(c^{\dagger}_{{\bf i}\sigma}c^{\phantom{\dagger}}_{{\bf j}\sigma} +
c^{\dagger}_{{\bf j}\sigma}c^{\phantom{\dagger}}_{{\bf i}\sigma} ) +
\sum\limits_{i\sigma} (\epsilon_{{\bf i}\sigma}
-\widetilde{\mu}_{{\bf i}\sigma}) n_{{\bf i}\sigma} \nonumber
\\ &+& \sum_{\bf i} [\Delta^{\phantom{\dagger}}_{\bf i}
c^{\dagger}_{{\bf i}\uparrow}c^{\dagger}_{{\bf i}\downarrow}
+\Delta^{*}_{\bf i} c^{\phantom{\dagger}}_{{\bf i}\downarrow}
c^{\phantom{\dagger}}_{{\bf i}\uparrow} ] \,\,
\label{Hbdg}
\end{eqnarray}
Here we have generalized the clean model to allow for site and spin
dependent local energies $\epsilon_{{\bf i}\sigma}$ and hoppings
$t_{{\bf ij}\sigma}$.  Eq.~\ref{Hbdg} also introduces a chemical
potential which includes the Hartree shift, $\widetilde{\mu}_{{\bf
i}\sigma}=\mu_{\sigma} + |U|\langle n_{{\bf i},-\sigma} \rangle$.
Randomness in the interactions can also be considered
\cite{aryanpour06} but we do not do that here.

The disorder we consider is uniformly distributed,
$P(\epsilon)=\frac{1}{2 V}$, $\epsilon\in[-V,V]$, and
$P(t)=\frac{1}{2 V_t}$, $t\in[t_0-V_t,t_0+V_t]$.
Evidently, $\langle \epsilon \rangle = 0$ and $\langle t \rangle = t_0$.
We scale all energy parameters $U$, $V$, $T$, $\Delta$, and $\mu$ to
units where $t_0=1$.  We restrict our study to either potential or
hopping disorders and do not analyze the case where both types of
disorder are present simultaneously.  We will find that the qualitative
features of the phase diagram is not very sensitive to the details of
the choice of disorder.

It is worth noting that the probability distribution of the experimental
disorder\cite{white09} is of exponential form
$P(\epsilon)=\frac{1}{V}e^{-\epsilon/V}$, and hence differs
from our bounded randomness \cite{foot1}. In addition, in experiments,
the fine-grain speckle is expected to disorder the hoppings,
interactions and site energies together.  For example, the tunneling
amplitude $t_{\bf ij}$ depends on the absolute difference of well-depths
of sites ${\bf i}$ and ${\bf j}$, and hence not only does it exhibit
randomness, but, in fact, randomness which is correlated with that in
the site energies.  Similarly, the interaction $U_{\bf i}$ depends on
$\epsilon_{\bf i}$ because the potential well modifies the
single-particle Wannier site basis functions, which determines $U_{\bf
i}$. It has been estimated \cite{zhou10} that the tunneling amplitude
disorder is characterized by a width $10^{-2}<\sigma_{t}<0.1$ relative
to its mean value, whereas $10^{-4}<\sigma_{U}<10^{-2}$, so that on-site
interaction can be taken constant\cite{zhou10}.

\subsection{Bogoliubov-de Gennes Treatment}

$H_{\rm eff}$, which is quadratic in the fermion operators, is
diagonalized via the Bogoliubov transformation,
\begin{eqnarray}
c^{\phantom{\dagger}}_{{\bf i}\uparrow} &=&
\sum_{n} \left [ \, \gamma^{\phantom{\dagger}}_{n\uparrow}
  u^{\phantom{*}}_{{\bf i}n} -
\gamma^{\dagger}_{n\downarrow} v^{*}_{{\bf i}n} \right ],
\nonumber \\
c^{\phantom{\dagger}}_{{\bf i}\downarrow} &=&
\sum_{n} \left [ \, \gamma^{\phantom{\dagger}}_{n\downarrow}
  u^{\phantom{*}}_{{\bf i}n} +
\gamma^{\dagger}_{n\uparrow} v^{*}_{{\bf i}n}\right ]. \,\,
\label{bdgtransform}
\end{eqnarray}
In the clean system the eigenfunctions $u_{n}$ and $v_{n}$ are plane
wave states.  In the presence of disorder they must be obtained by
(numerical) diagonalization.  The local order parameter and density
are determined self-consistently,
\begin{eqnarray}
\Delta_{\bf i}=-|U|\langle c^{\phantom{\dagger}}_{{\bf i}\downarrow}
c^{\phantom{\dagger}}_{{\bf i}\uparrow} \rangle
&=& -|U| \sum_n f(E_n)  u^{\phantom{*}}_{{\bf i}n} v^{*}_{{\bf i}n},
\nonumber \\
\langle n^{\phantom{\dagger}}_{{\bf i}\uparrow} \rangle &=& \sum_n
f(E_n) | u^{\phantom{\dagger}}_{{\bf i}n}|^2,
\nonumber \\
\langle n^{\phantom{\dagger}}_{{\bf i}\downarrow} \rangle &=& \sum_n
f(-E_n) | v^{\phantom{\dagger}}_{{\bf i}n}|^2,
\,\,
\label{selfconsistent}
\end{eqnarray}
where $f$ is the Fermi function. These self-consistency conditions
are equivalent to minimizing the free energy. To start the process of
solving the BdG Hamiltonian self-consistently, an initial random guess
for $\{\Delta_{\bf i}\}$ and $\{\langle n_{{\bf i},\sigma} \rangle\}$ is
made at every site. This guess is inserted into the matrix form of
$H_{\rm eff}$ and the matrix diagonalized to get the BdG eigenvalues
$\{E_n\}$ and eigenvectors $\{u_n,v_n\}$. From these eigen-pairs, a new
set of $\{\Delta_{\bf i}\}$ and $\{\langle n_{{\bf i},\sigma} \rangle\}$
is computed using Eq.~\ref{selfconsistent}. This new set is reinserted
into $H_{\rm eff}$ and the Hamiltonian matrix rediagonalized, with the
resulting eigen-pairs again fed into Eq.~\ref{selfconsistent}. Each
cycle of this process defines one iteration, and the iterations are
repeated until the $\{\Delta_{\bf i}\}$, $\{\langle n_{{\bf i},\sigma}
\rangle\}$ at every site differ from those of the previous iteration to
within the specified accuracy, i.e, self-consistency is attained at
every site\cite{ghosal98,ghosal01}. The chemical potentials are also
adjusted in each iteration to achieve the desired density to the same
accuracy. This accuracy is $10^{-5}$ in all our final, self-consistent
results.

To study the phase diagram, we define a spatially averaged order
parameter, $\Delta_{\rm op}$, from $\Delta_{\bf i}$.  The BdG spectrum
can be used to determine the energy gap, $E_{\rm gap}$, which is
the lowest eigenvalue above the chemical potential.  The spectrum and
eigenfunctions also determine the density of states.  Note that unlike
the spin-independent case the eigenvalues do not come in $\pm$ pairs and
the distances of closest eigenvalues below and above the chemical
potential are in general different.  In principle, these latter two
eigenvalues can be used to examine separately the positive gap (the
energy cost to add an additional quasiparticle to the system), and the
negative gap (the energy cost to extract a quasiparticle from the
system, or to create a quasihole).  In practice, however, we find their
magnitudes are always approximately equal, so we show the data only for
the positive gap.

In generating the phase diagrams, we average all our data over $5-10$
disorder realizations so that our results are not characteristic of
any particular realization.  Such disorder averaging restores some of
the symmetries of the model, as discussed in the following subsection.
The greatest variation about the mean of results for individual
realizations occurs close to the critical points, which is a typical
signature of phase transition. Away from the critical point we find
that the values of $E_{\rm gap}$ and $\Delta_{\rm op}$ vary only by a
few percent from realization to realization on lattices of size 24$\times$24.

\subsection{Symmetries of the Model}

Symmetries of the Hamiltonian allow us to identify the conserved
quantities (and degeneracies of states), which in turn can simplify
finding, and interpreting, the solutions.  In considering the
symmetries of Eq.~\ref{Hbdg}, it is useful to note that the
spin-dependent term of $H_{\rm eff}$ can be written as the sum of a
random local chemical potential $ \frac{1}{2} (\epsilon_{{\bf
i}\uparrow} + \epsilon_{{\bf i}\downarrow} ) \, (n_{{\bf i}\uparrow} +
n_{{\bf i}\downarrow}) $ and a random local Zeeman field, $\frac{1}{2}
(\epsilon_{{\bf i}\uparrow} - \epsilon_{{\bf i}\downarrow} ) \,
(n_{{\bf i}\uparrow} - n_{{\bf i}\downarrow})$. This allows us to make
connections to previous work in which Zeeman field terms are
considered, as we point out in our results section.


The most obvious implication of disorder on symmetries of the model is
that it breaks translational invariance.  In fact, it is often useful
to take advantage of the spatial inhomogeneities in individual
realizations of randomness, for example, by examining correlations
between the distribution of the local $\Delta_{\bf i}$ and excited state wave
functions.  (See below and the discussion in
Ref.~\onlinecite{ghosal01}).  Disorder averaging restores
translational symmetry, at least for physical quantities like
correlation functions.  Most of our results reflect this
averaging.


The spin-dependent terms break spin rotational invariance and have
implications for time-reversal symmetry.  If we define $T$ as the
second-quantized time-reversal operator, we find that $[T,H_{\rm
eff}]=0$ provided that $\Sigma_{\bf i} \epsilon_{{\bf i}\sigma}=0$.
This means time reversal symmetry is preserved in the mean field
approximation, provided that we also choose
$\mu_{\uparrow}=\mu_{\downarrow}$.  As discussed in Ref.~\onlinecite{jiang11}, when
there is site disorder only on one spin species, the chemical
potentials required to maintain equal populations are indeed identical
for disorder strengths below a critical threshold.  However, beyond
that value, equal populations occur only when the chemical potentials
are tuned to different values.  In that situation, time reversal
symmetry is broken spontaneously, $[T,H_{\rm eff}]\neq 0$.

Time reversal symmetry implies that every energy eigenstate is at
least doubly degenerate. Anderson's theorem states that for weakly
disordered superconductivity, noninteracting eigenstates related by
time reversal can be paired with each other to form Cooper
pairs\cite{anderson59}. When time reversal is broken at $V_{c}^{\rm
gap}$ in our BdG model, there are no more pair states related by time
reversal so the tendency for SC order is diminished. This is the
qualitative origin of the transitions we observe which are absent in
the spin-independent case.

There can still be pairing between time reversal broken pairs. As an
example, in 2D clean systems with a uniform parallel Zeeman field, one
has the pairing term $\langle
c_{\mathbf{k}\uparrow}c_{-\mathbf{k}+\mathbf{q}\downarrow} \rangle$,
in addition to the spin population imbalance. This is the
Fulde-Ferrell-Larkin-Ovchinnikov (FFLO)
condensate\cite{fulde64,larkin64} which carries a finite momentum
${\bf q}$, and when expressed as a stationary wave, has sinusoidal,
sign-changing $\Delta_{\bf i}$. Similarly, when we relax the
constraint of spin balance (but keep total density fixed with a single
$\mu$) in our system, we find that the spin population becomes
imbalanced beyond the $\mu_{\sigma}$ bifurcation point, and the
$\Delta_{\bf i}$ has regions of sign-changing values. We thus obtain a
transition from disordered BCS to disordered FFLO (or dLO) when we
relax density conservation of each spin (ie: fixed $\langle n_{\sigma}
\rangle$). The time reversal non-invariance of $H_{\rm eff}$ means the
ground-state is not time reversal invariant as well, implying that it
carries finite momentum, and hence must have a spatially varying phase
$e^{i\mathbf{q}\cdot\mathbf{r}}$.


Finally, we consider particle-hole symmetry.
As is well known, on bipartite lattices the kinetic and
interaction terms are invariant under the transformation,
\begin{eqnarray}
c_{{\bf i}\sigma}^{\phantom{\dagger}}&=&
(-1)^{{\bf i}}d^{\dagger}_{{\bf i}\sigma},
\nonumber \\
c^{\dagger}_{{\bf i}\sigma}c_{{\bf i}\sigma}^{\phantom{\dagger}}&=&
1-d^{\dagger}_{{\bf i}\sigma} d_{{\bf i}\sigma}^{\phantom{\dagger}}.
\label{particlehole}
\end{eqnarray}
Here the phase factor $(-1)^{\bf i} = +1(-1)$ on the A(B) sublattices
respectively.  This is true even when the hopping and interaction
strengths are random.

Local chemical potential terms are however not invariant under this
particle-hole transformations, but instead change sign (and introduce
an irrelevant constant shift to the Hamiltonian).  This seemingly
suggests that the two models with random site energies $\epsilon_{{\bf
i}\sigma}$ are not particle-hole symmetric.  However, because the
distribution of randomness is chosen so that
$P(\epsilon)=P(-\epsilon)$, particle-hole symmetry is satisfied on
average.  Meanwhile, the third model with random hopping $t_{\langle
{\bf i}{\bf j} \rangle\sigma}$ is exactly particle-hole symmetric.

The consequence of these observations is that all correlation
functions and the resulting phase diagram of the random hopping
Hamiltonian are precisely symmetric about half-filling.  Approximate
symmetry is expected, and confirmed numerically for site energy
disorder.  In what follows we therefore show results only for $n \leq
1$.  We note that the presence or absence of particle hole symmetry
has been found to be central to the appearance of metal-insulator
transitions in the repulsive Hubbard Hamiltonian \cite{denteneer01}.

\section{Common Properties of Spin-Dependent Disorder}

In this section we describe properties shared by all three
spin-dependent disorder Hamiltonians. Although of course the
quantitative positions of the boundaries vary, the basic topology of
the phase diagrams is the same.
We use the following
notation to refer to the models: $V_{\uparrow\downarrow}$ refers to
the model with site disorder on both spins, $V_{\uparrow}$ to the
model with site disorder on one spin only, and
$t_{\uparrow\downarrow}$ to the model with hopping disorder on both
spins.

In constructing the phase diagrams, we define
the vanishing of $\Delta_{\rm op}$ and $E_{\rm gap}$ to occur when
their values become less than $10^{-2}t$.
Above the critical point,
$E_{\rm gap}$ and $\Delta_{\rm op}$
fluctuate randomly on the {\it same} order as the $0.01t$ cutoff.
We have also shown that these residual values scale to zero
with increasing lattice size.\cite{jiang11}
A final justification for our chosen cutoff comes from the
observation that the
realization-to-realization fluctuations of $E_{\rm
gap}$ and $\Delta_{\rm op}$ are comparable to the cutoff.


We also use the terms ``No SC" and ``(Anderson) Insulator"
interchangeably since, quite generally, the unordered phase of a
non-interacting 2D lattice is an Anderson Insulator for arbitrary
disorder strength. The presence of attractive interactions is expected
to enhance further the effective depth of the potential wells (minima of
$\epsilon_{\bf i \sigma}$), thus making the localization tendency for
the opposite spin-species greater in the mean-field approximation.

All phase diagrams are for $U=-2$ and $n=0.875$ unless we are varying
$|U|$ or $n$. Lattice size of 24$\times$24 is used throughout.

\subsection{Phase Diagram in Disorder-Temperature plane}

\begin{figure}
 \includegraphics[width=0.4\textwidth,angle=-90]{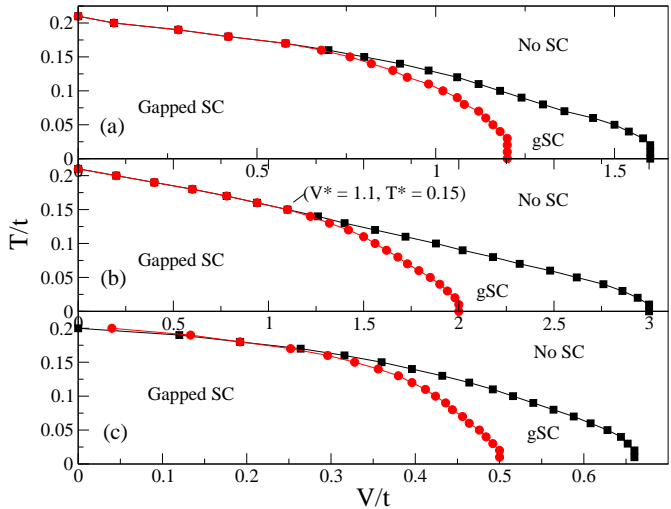}
 \caption{$V$-$T$ phase diagrams for $U=-2$, $n=0.875$ for (a)
 $V_{\uparrow\downarrow}$, (b) $V_{\uparrow}$, (c)
 $t_{\uparrow\downarrow}$. All three panels show the phase-boundaries
 for $\Delta_{\rm op}$ (black squares) and $E_{\rm gap}$ (red
 circles). Horizontal axes for the three panels are not to the same
 scale. ($V^{*}$, $T^{*}$) marks the critical point where the gSC
 phase vanishes.
 \label{fig1} }
\end{figure}

Figure \ref{fig1} shows the $V$-$T$ phase planes for the models (a)
$V_{\uparrow\downarrow}$, (b) $V_{\uparrow}$, and (c)
$t_{\uparrow\downarrow}$. In Fig.~\ref{fig1} (b), for example, the
phase boundary $T_{c}(V)$ is the common line for vanishing $E_{\rm
gap}$ and $\Delta_{\rm op}$ for weak disorder strength $0\leq V_{c}
\lesssim 1.2$. At the critical point ($V^{*}$=1.1, $T^{*}$=0.15), the boundary
bifurcates into two curves $T_{c}^{\rm gap}(V)$ and $T_{c}^{\rm
op}(V)$ with $T_{c}^{\rm gap}(V)$ $<$ $T_{c}^{\rm op}(V)$. The region
enclosed by the $V$, $T$ axes and $T_{c}^{\rm gap}(V)$ is the gapped
SC phase. The region between the two curves is the gapless SC (gSC)
phase in which there is coexistence of Cooper paired and unpaired
fermions. The excitations in this region cost no energy because their
low-energy single-particle MF wave-functions (weight or amplitude
squared) have no overlap with the regions of significant bound pairs,
i.e., large $\Delta_{\bf i}$. We provide evidence for this in Section~\ref{sec:Disc} below.

\begin{figure}[!]
\includegraphics[width=0.4\textwidth,angle=-90]{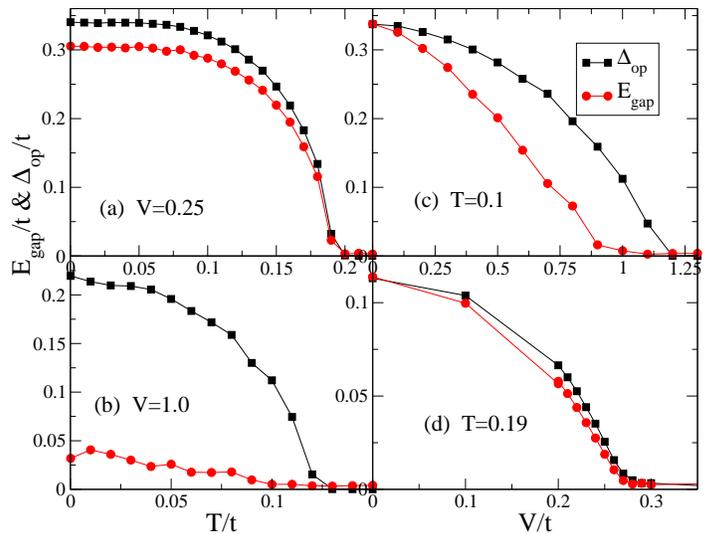}
 \caption{ $E_{\rm gap}$, $\Delta_{\rm op}$ vs $T$ for (a) weak disorder
 $V=0.25$ showing simultaneous transitions for both $E_{\rm gap}$ and
 $\Delta_{\rm op}$ at $T_{c}=0.19$; (b) strong disorder $V=1$: The
 transition of $E_{\rm gap}$ takes place first at $T_{c}^{\rm
 gap}=0.1$, followed by that of $\Delta_{\rm op}$ at $T_{c}^{\rm
 op}=0.12$. $E_{\rm gap}$, $\Delta_{\rm op}$ vs $V$ for (c)
 intermediate temperature $T=0.1$, with the critical points
 $V_{c}^{\rm gap}=0.9$, $V_{c}^{\rm op}=1.2$; (d) high temperature
 $T=0.19$ with a single common critical point $V_{c}=0.27$. All panels
 are for the $V_{\uparrow\downarrow}$ model with $U=-2$, $n=0.875$. We
 identify the critical points by using the criterion that a quantity
 vanishes when its value becomes less than $10^{-2}$. \label{fig2} }
\end{figure}

In Figure~\ref{fig2} (a,b,c,d) we show $E_{\rm gap}$ and $\Delta_{\rm
op}$ as functions of $T$ (a,b) or $V$ (c,d) for
$V_{\uparrow\downarrow}$. These are the representative cuts through
the $V$-$T$ diagram of Fig.~\ref{fig1} (a). The phase diagram,
Fig.~\ref{fig1}, is generated by sweeping disorder strength at fixed
temperature, or sweeping temperature at fixed disorder.
Figure~\ref{fig2} (a,b) shows two such vertical cuts: one at weak
disorder for which there is no gapless SC phase, and one at larger
disorder for which there is a gSC phase.  Similarly, Figure~\ref{fig2}
(c,d) shows horizontal cuts at fixed temperature: one at an
intermediate temperature for which again there is a gSC phase, and one
for a high temperature for which there is no gSC phase.
Another feature to note
in Figs.~\ref{fig2} (b,d) is that in the gSC phase $\Delta_{\rm op}$
(and $E_{\rm gap}$) is substantially reduced due to the presence of
broken pairs, as noted already.

As a cross-check, we note that that the transition points have
consistent values between Figs.~\ref{fig2} (a) and ~\ref{fig2} (d), as
well as between Figs.~\ref{fig2} (b) and ~\ref{fig2} (c).
For stronger coupling $U$, the gapless SC region at $T=0$ is larger and
consequently covers larger areas of the $V$-$T$ phase plane. Its closing
point moves to larger values of both ($V^{*}$, $T^{*}$).

\subsection{Phase Diagram in Interaction-Disorder plane}

\begin{figure}
\includegraphics[width=0.4\textwidth,angle=-90]{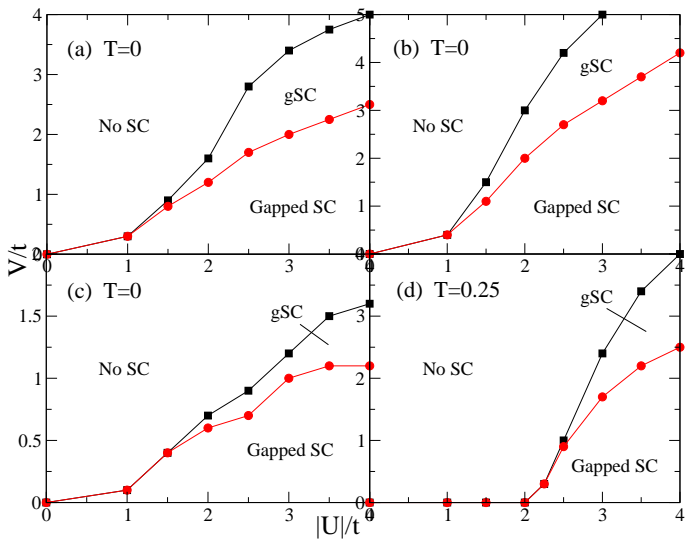}
 \caption{$U$-$V$ phase diagrams for $T=0$ (low-$T$), $n=0.875$ for (a)
 $V_{\uparrow\downarrow}$, (b) $V_{\uparrow}$, (c)
 $t_{\uparrow\downarrow}$; (d) $T=0.25$ (high-$T$), $n=0.875$ for
 $V_{\uparrow\downarrow}$. In (a,b,c), because of Anderson
 localization for arbitrarily small disorder strength and BCS pairing
 for arbitrarily small $|U|$, the apex of the gSC phase must actually
 coincide with the origin. In (d), the SC phases are destroyed for
 larger $|U|$ values at finite $T$. By comparing (a) and (d), we see
 that finite $T$ affects phase-boundaries in the weak-$V$ region of
 the plane more than the strong-$V$ region. \label{fig3} }
\end{figure}

We show the $U$-$V$ phase diagrams for $T=0$ and $n=0.875$ in
Figs.~\ref{fig3} (a,b,c) for the three models.  In the non-interacting
limit ($U=0$), the entire vertical axis is an Anderson insulating phase.
In the clean limit ($V=0$), the entire horizontal axis is
covered by the BCS SC phase.
For non-zero $U$ and $V$,
we find three distinct phases:
[1] The Anderson
insulator bordered by $V$ axis, the origin, and a positive sloped
curve passing through the origin; [2] The gapped SC phase bordered by
the $|U|$ axis, origin, and a less sloped curve; and [3] the gSC phase
sandwiched in between.\cite{footnote1}

The $U$-$V$ phase diagram at finite temperature, $T=0.25$, is shown in
Fig.~\ref{fig3} (d) for $V_{\uparrow\downarrow}$.
As expected, the SC phases shrink due to thermal fluctuations.
A finite $U_c$ is now required to get SC even in the clean limit.
It also appears that while the gSC intervenes between the SC and
Anderson insulator at $T=0$ when $|U| \sim 2$, there is a direct
transition at finite $T$ for this interaction strength.  We attribute
the diminished role of disorder in giving rise to the gSC phase to the
smoothing effect of thermal effects on the random potential.



\subsection{Phase Diagram in Interaction-Temperature plane}

\begin{figure}
\includegraphics[width=0.4\textwidth,angle=-90]{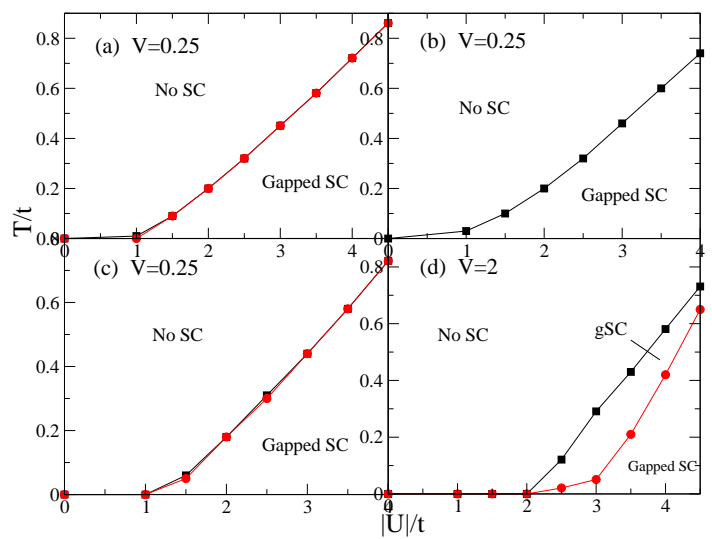}
 \caption{$U$-$T$ phase diagrams for $V=0.25$ (low-$V$), $n=0.875$ for (a)
 $V_{\uparrow\downarrow}$, (b) $V_{\uparrow}$, (c)
 $t_{\uparrow\downarrow}$; (d) $V=2$ (high-$V$), $n=0.875$ for
 $V_{\uparrow\downarrow}$. In (a,b,c), at low-$V$, there is no
 intervening gSC phase and there is a direct transition from gapped SC
 to insulator in the $|U|$-$T$ plane. In (d), at strong-$V$, a gSC
 region opens up in the $|U|$-$T$ plane. \label{fig4} }
\end{figure}

We show the $U$-$T$ phase diagrams
in Fig.~\ref{fig4} (a,b,c) for all three models at $V=0.25$ and for
stronger disorder $V=2$ for the $V_{\uparrow\downarrow}$ model
in Fig.~\ref{fig4} (d). The critical lines $T_{c}^{\rm
gap}(|U|)$ and $T_{c}^{\rm op}(|U|)$ show a nearly linear increase
with $|U|$, and, for weak disorder, coincide. In Fig.~\ref{fig4} (d), there is a
gSC region for  $(0<T<0.8, \,\,\, 2<|U|<5)$.
These results emphasize that the gSC phase is driven by the disorder $V$.

\subsection{Phase Diagram in Filling-Disorder plane}

In Figs.~\ref{fig6} (a,b) we display the phase diagrams of the
$V_{\uparrow\downarrow}$ model for $U=-2$ in the filling-disorder
plane for $T=0$ (low-temperature) and $T=0.1$ (high-temperature)
respectively. We restrict $n$ to the interval $0 \leq n \leq 1$
because of approximate particle-hole symmetry of the potential
disorder models.
At $T=0$,
Fig.~\ref{fig6} (a) shows that as the disorder strength is increased
there are transitions from gapped SC to gSC to insulator for all
filling values except the trivial $n=0$ case. However, at a higher
temperature $T=0.1$ the SC order is destroyed for filling
values below a critical filling $n_{c}=0.4$ for all $V$
(Fig.~\ref{fig6} (b)). For fillings greater than $n_{c}\approx 0.6$
a gSC phase appears.
At low density, the effects of disorder are often larger
because only the lowest energy levels, with greatest deviation
from the mean, are occupied.

\subsection{Phase Diagram in Filling-Temperature plane}

\begin{figure}
\includegraphics[width=0.4\textwidth,angle=-90]{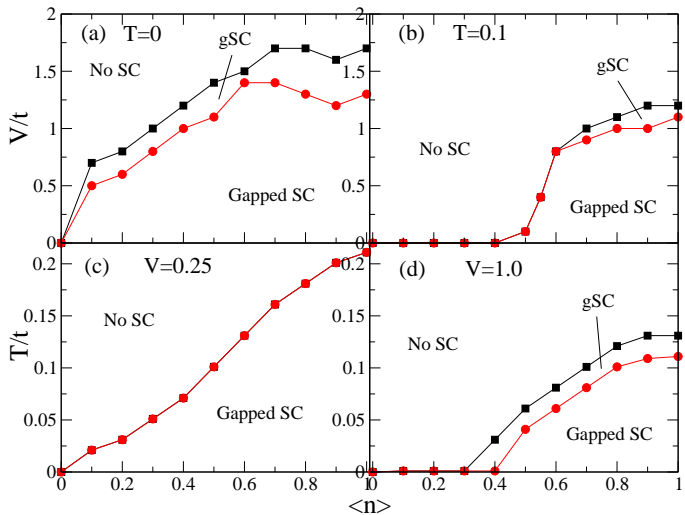}
 \caption{(a), (b) Phase diagrams in the filling-disorder plane for
 interaction $U=-2$: (a) $T=0$ (low-$T$), (b) $T=0.1$ (high-$T$). (c), (d)
 Phase diagrams in the filling-temperature plane: (c) $V=0.25$
 (low-disorder), (d) $V=1.0$ (high-disorder). At high-$T$ (b), the
 locations of the phase-boundaries are reduced and there exists a
 critical filling $n_{c}\neq 0$, compared to low-$T$ in (a). Likewise,
 in (c) and (d), as we go from low-$V$ to high-$V$. There is no gSC
 phase for low-$V$ in (c), but there is one at high-$V$ in (d). All
 panels are for the $V_{\uparrow\downarrow}$ model, and shown only in
 the interval $0\leq n \leq 1$. \label{fig6} }
\end{figure}

The phase diagram in the filling-temperature plane, again for $U=-2$,
is displayed in Figures~\ref{fig6} (c,d) for (c) low disorder,
$V=0.25$, and (d) higher disorder, $V=1.0$. In Fig.~\ref{fig6} (c), at
low disorder, there is a single line of transitions from the gapped SC
phase to the insulator phase, with no intervening gSC phase.
When the disorder
is high, Fig.~\ref{fig6} (d) shows that at zero temperature,
gapped SC order occurs for $n_{c} \geq 0.4$, and a gSC phase for $0.3 \leq
n_{c} \leq 0.4$.

At zero disorder, it is known that $T_{c}=0$ for the square lattice
attractive Hubbard at half-filling ($n=1$) due to the degeneracy
between superconducting and charge density correlations.
It is worth noting, however, that away
from half-filling the transition temperature scale is not that
dramatically different from what is found in QMC and other non-MF
approaches,\cite{denteneer94,paiva04} where $T_{c} \approx 0.1t$.
It is worth noting another well-known difficulty with MF approaches,
namely that the SC transition increases with $U$ at strong coupling
(see Figs.~\ref{fig4},~\ref{fig6}) whereas the exact $T_c$ turns over and falls with
$1/U$.

\section{\label{sec:Disc}Discussion of the Common Properties}

In describing the effects of disorder, one often makes the general
argument that the specific form of disorder is irrelevant to transitions,
since under a renormalization flow, disorder in one term will
propagate into others.  This would suggest that, as we have found,
the phase diagrams of our models corresponding to different
types of randomness, should be qualitatively similar.
However, this argument is not immediately compelling because
in some cases the disorder has different symmetry properties
\cite{denteneer01}.  Apparently, this does not occur here.
Here we make some brief observations on the quantitative
differences between the phase diagrams.


For a fixed set of model parameters, particularly disorder strength,
the values of $E_{\rm gap}$ and $\Delta_{\rm op}$ are higher for the
$V_{\uparrow}$ site disorder than for $V_{\uparrow\downarrow}$ site
disorder. Also, the values of $T_{c}$ and $V_{c}$ for destroying SC
are higher for the former model compared to the latter.
Similar to the $V_{\uparrow\downarrow}$ model,
the $t_{\langle {\bf i}{\bf j} \rangle,\sigma}$ of the hopping
disorder are uncorrelated on the same bond for different $\sigma$, so
the vanishing of $E_{\rm gap}$ and $\Delta_{\rm op}$ take place at
lower $T_{c}$ and $V_{c}$ as compared to $V_{\uparrow}$ model. See
Figure~\ref{fig1} for a quantitative comparison of the how $T_{c}$ and
$V_{c}$ differ between each of the three models.

\begin{figure}
\centering
\includegraphics[width=0.5\textwidth,angle=0]{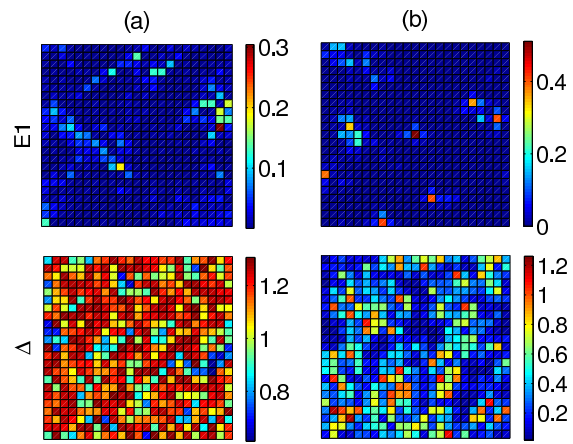}
\caption{Real-space plots for the $V_{\uparrow\downarrow}$ model. Left panels (a): Sum of wave-function amplitude-squared for
the first five excited states above and below `$\mu$' and $\Delta_{\bf
i}$ at $T=0.4$ for $U=-4$, $V=0.5$ (in the gapped SC phase). These
show the overlap of the weight of the single-particle low-energy
excitations with regions of significant SC order. Right panels (b):
The same for $U=-4$, $V=2.0$ (gSC phase), which show the non-overlap
of the low-energy states with the SC islands, leading to the gapless
SC phase. \label{fig7} }
\end{figure}

A common qualitative feature of the phase diagrams is the appearance
of a gSC phase.  This occurs because in all these models it
is possible for the low lying excited states to be located
in regions where the superconducting order parameter vanishes
(even though there are also significant superconducting regions).
To illustrate this mechanism, first discussed in
[\onlinecite{ghosal01}],
we have correlated the regions of
high or low $\Delta_{\bf i}$ with the probability of finding
quasi-particles in the lowest (highest) five excited states.
Here we use only
a single disorder realization of the $V_{\uparrow\downarrow}$ model.
Figure~\ref{fig7} (a) shows
significant overlap between $\Delta_{\bf i}$ and the highest
probability for quasi-particle/hole excitations in the ten states,
while Figure~\ref{fig7} (b) shows the non-overlap of these
regions. This leads to finite $T$ gapped and gSC excitations
respectively.

The presence of a uniform Zeeman field on a clean (disordered)
attractive Hubbard lattice is known to induce a spin-imbalance and
resulting FFLO (dLO) phases with sign-changing energy gap $\Delta_{\bf
i}$ within BdG MFT\cite{cui08}. Since we have random Zeeman fields in
two of our models (and hence local spin-imbalance), it is natural to
ask whether sign changes in $\Delta_{\bf i}$ occur.  We find that our
models do not have a sign changing $\Delta_{\bf i}$, and, in fact, that such sign
changes are a diagnostic that we have converged to an incorrect,
metastable solution.

This result is consistent with a recent BdG study of a 1D balanced
system $N_{\uparrow}=N_{\downarrow}$ which similarly found no
sign-changes in $\Delta_{\bf i}$ in the presence of a single magnetic
impurity, despite the presence of local spin-imbalance and a reduced
value of $\Delta_{\bf i}$ at and around the defect \cite{ljiang11}.
On the other hand, in the same study, the authors also find that a
sufficiently extended magnetic impurity in 1D and 3D, for example,
with a Gaussian profile, does result in both a local spin-imbalance
and sign-changes in $\Delta_{\bf i}$ (FFLO phase).  Similarly, another
recent BdG study\cite{zapata10} in 1D with
$N_{\uparrow}=N_{\downarrow}$ and a spin-dependent lattice
[$V_{\uparrow}(x)=-V_{\downarrow}(x)=V_{0}{\rm cos}(2 \pi x/\lambda)$]
concluded that ``$\pi$ phases" which exhibit sign-changing
$\Delta_{\bf i}$ require spin-dependent lattices with wave-length
longer than the coherence length.

Thus the absence of sign-changing $\Delta_{\bf i}$ in our work is due
to our choice of spatially uncorrelated random magnetic impurities at
every site and consequently a spin-dependent lattice potential with
wavelength much less than the coherence length.  Current experiments
can introduce such very rapidly varying laser speckle
\cite{aspect09,palencia10}.


\section{\label{sec:Spec}Special Cases of the Models}
To understand better the physics of spin-dependent disorder within the 2D attractive Hubbard model, we explored some particular features of the $V_{\uparrow\downarrow}$ model.

\begin{figure}[!]
\includegraphics[width=0.35\textwidth,angle=-90]{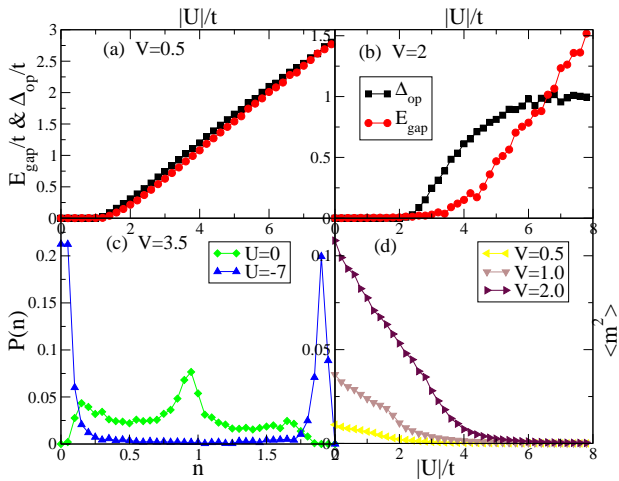}
 \caption{For the $V_{\uparrow\downarrow}$ model, $E_{\rm gap}$, $\Delta_{\rm op}$ vs $|U|$ for (a) $V=0.5$:
 $E_{\rm gap}$ and $\Delta_{\rm op}$ increase without bound at low
 disorder; (b) $V=2.0$: at high disorder, $E_{\rm gap}$ increases
 without bound while $\Delta_{\rm op}$ saturates. (c) The particle
 distribution $P(n)$ for $V=3.5$, $U=0$: There is roughly a three-peak
 structure implying that most sites are empty, singly-, or
 doubly-occupied for spin-dependent disorder at weak interactions. For
 the same $V$ and $U=-7$, the peak near $n\approx 1$ has disappeared,
 so single occupancy is disfavored for $|U|\gg V$. (d) The local-moment
 $\langle m^{2} \rangle$ vs $|U|$ for $V= 0.5$, 1.0, 2.0: As $V$
 becomes stronger, the initial local-moment is higher. However, as
 $|U|$ is increased, the local-moment is driven to zero in the regime
 $|U|\gg V$. All panels for $T=0$, $n=0.875$. \label{fig10} }
\end{figure}

In Figure~\ref{fig10} (c) we show the particle density histogram for
$|U|=0,7$.  For $|U|=0$, we find the disorder landscape favors site
fillings of approximately zero, one, or two particles. For $|U|=7$, we
see that the histogram has only two peaks at fillings of zero and two.
This can also be seen in Figures~\ref{fig10}
(a,b). Fig.~\ref{fig10} (a), for $V=0.5$, shows $E_{\rm gap}$ and
$\Delta_{\rm op}$ increase without bound versus $|U|$ and $\Delta_{\rm op}
\geq E_{\rm gap}$, as is always the case for spin-dependent disorder;
but in Fig.~\ref{fig10} (b), for $V=2.0$ and for strong enough $|U| \gg
V$, we find that $\Delta_{\rm op} \leq E_{\rm gap}$ and saturates. In
previous spin-independent BdG work in
Ref. \onlinecite{ghosal01,ghosal98}, it was found that $\Delta_{\rm
op} \leq E_{\rm gap}$ for all disorder strengths. Here we find that we
recover some of the results of the spin-independent disorder for $|U|
\gg V$.

In Fig.~\ref{fig10} (d) we show the evolution of the site-averaged,
squared local-moment, $\langle m^{2} \rangle$, as a function of $|U|$
for three disorder strengths, $V=0.5, 1.0, 2.0$ at $T=0$. When
$|U|=0$, the moment starts out at its maximum possible value. This
maximum value is larger for greater $V$ because the potential minima
for the $\uparrow$-fermions and $\downarrow$-fermions are in general
at different sites due to the spin-dependent randomness. As $|U|$
increases, it becomes energetically favorable for two fermions of
opposite spin to occupy the same site forming a singlet, and as a
result, decreasing the local moment all the way to zero for very large
$|U|$.

The destruction of the average local-moment is seen to occur at
roughly the same $|U|$ value where the cross-over from $\Delta_{\rm
op} > E_{\rm gap}$ to $\Delta_{\rm op} < E_{\rm gap}$ takes place in
Fig.~\ref{fig10} (b). Spin-independent disorder always has the
property $\langle m^{2} \rangle = 0$, since both spin-species see the
same potential landscape. The average local-moment destruction for
large $|U|$ is, therefore, also an indication that in the $|U| \gg V$
limit, we are in the spin-independent regime.

\section{Conclusions}


Before summarizing the results of this paper, we relate our study to
previous classic investigations of magnetic impurities in
superconductors.
One of the earliest works on this topic was by Abrikosov and Gor'kov
(AG)\cite{ag61} who considered a continuum model of a weak-coupling
superconductor with dilute magnetic impurities in the form of randomly
distributed and oriented classical spins coupled to the electrons via a
rotationally invariant exchange interaction.  As with our Zeeman
interaction, this external magnetic field breaks time-reversal
invariance and suppresses superconductivity since it is not protected by
Anderson's Theorem \cite{anderson59}.  AG's mean field treatment uses
the perturbative, self-consistent, Born approximation.
Although the magnetic
disorder results in an inhomogeneous $\Delta_{\bf i}$, due to
the diluteness of the impurities, $\Delta_{\bf i}$ is rather different
from that considered here.  It is predominantly spatially uniform
except in the vicinity of the impurity moments.  In fact,
although their formalism admits inhomogeneity,
in solving the resulting integral equation the
order parameter is assumed constant.  AG find that the
quasiparticle energies and $\Delta_{\rm op}$ are modified in different
ways and, as a consequence $E_{\rm gap}$ is suppressed more than
$\Delta_{\rm op}$.  The more rapid decay of $E_{\rm gap}$  leads to a
gapless SC phase, as found here.


To check the validity of AG theory in the strong disorder regime, we modified our BdG codes to
enforce a spatially uniform $\Delta_{\bf i}$ equal to the site average of the inhomogeneous $\Delta_{\bf i}$.
We found that the region of gapless SC phase is greatly diminished compared to the case of inhomogeneous
$\Delta_{\bf i}$ \cite{jiang11}. This illustrates the limitations of AG theory in addressing the physics
of strong disorder, and it also validates our physical picture for the gapless SC phase.
A nearly uniform $\Delta_{\bf i}$ precludes the possibility of a mechanism for the gapless SC
phase based on our physical picture (see below).


Like AG, our work is at the mean-field level. However, since it is based on
the non-perturbative BdG MFT, it can capture strong coupling effects, as
well as strong inhomogeneities in $\Delta_{\bf i}$ which occur
throughout the lattice and not just on widely separated impurity sites.
As described earlier, these effects lead to a clear physical picture for
the gapless SC phase namely, that the lowest lying quasi-particle/hole
excited states lie in the Anderson insulator regions where there is no
SC order, and hence a zero excitation gap. Such an interpretation is
absent from AG theory.  A further distinct feature of our work, as
compared to the AG treatment, is that the non-interacting eigenstates
are Anderson localized (in regions which are different for the up and
down spins) so our `normal' state is a (spin-dependent) Anderson
Insulator, whereas the normal state of the AG model is band metallic.


In this paper we have explored the interplay of strong attractive
interactions and spin-dependent disorder with rapid spatial variation.
We have mapped out the ground state phase diagrams
for various cases- when one species moves
in an environment completely decoupled from the disorder, when both
species see randomness, but the energy landscape is species dependent,
and finally when the hopping is
random, but dependent on the fermion spin.
We find that, as the disorder is increased, in some
situations the superconducting gap goes to zero first, followed by the
order parameter.  That is, there can be a gapless superconducting
phase.


Our numerical results are for systems of finite size,
most typically 24$\times$24 lattices.  Some finite size scaling was used, for
example to demonstrate small residual values of $E_{\rm gap}$ scale
to zero with increasing system size.
We should mention, however, that on these finite lattices
it is difficult to access any physics associated with rare regions,
e.g.~an exponential `tail' density-of-states
which starts from the gap edge ($E_{\rm gap}$) at a maximum and extends
to $E=0$, the so-called ``soft gap"
\cite{balatsky97,lamacraft00,lamacraft01,balatskyRMP06}.


We have considered the phase diagram/equilibrium thermodynamics here,
but transport properties pose some interesting questions.  For
example, when $U=0$ and one spin species sees no disorder, it will be
metallic while the other is insulating.  As the attractive $U$ is
turned on, the metallic species will see an induced randomness.  At
$U$ nonzero will both species be localized or extended, or can there
be parameter regimes where there is a `spin selective Anderson
transition'?

Mean field approaches such as the BdG treatment used here are, of
course, only approximate, and are problematic at finite temperatures
since they ignore phase fluctuations.  Studies of the attractive
Hubbard model with Quantum Monte Carlo methods offer a way to treat
the interactions exactly, although on lattices of smaller
size than with the BdG approach described here.  Quantum Monte Carlo
can go to very low temperatures and access the superconducting phase,
if it exists, because there is no sign problem either in the clean
case \cite{scalettar89,moreo91,imada91,assaad94} or for randomness
which is the same for both spin species \cite{huscroft97,hurt05}.
Unfortunately, spin-dependent disorder will introduce a sign problem
\cite{loh90} in determinant Quantum Monte Carlo even for the case of
attractive interactions.  Indeed, once one allows for both a non-zero
chemical potential and a non-zero Zeeman field, the positive and
negative $U$ Hubbard models can be mapped onto each other exactly.
Thus we expect limited ability to address the physics of spin
dependent disorder with this method.  Studies via dynamical mean field
theory, on the other hand are feasible at low temperature, and would
offer a more precise theoretical point of contact than the present
mean field treatment with optical lattice experiments when they are
performed.

\section{Acknowledgements}

This work was supported by the Department of Energy under
DE-FG52-09NA29464; by the National Science Foundation under grants
PHY-1005503 and PHY-1004848 and by the CNRS-UC Davis EPOCAL LIA joint
research grant.


\begin{thebibliography}{80}
\bibitem{foot1}
In some situations, in the thermodynamic limit, unbounded disorder
can produce somewhat different effects from bounded disorder.
\bibitem{footnote1}
We note that in MF calculations in the clean limit at $T=0$, finite
lattice effects obscure the BCS phase at small $|U|$.  Simulation of
larger lattices is required to get the precise boundaries in the region
$U<t$ of Figure 3.
\end{thebibliography}
\end{document}